\documentclass{ws-procs9x6-cpt25}

\usepackage{xspace}
\usepackage{relsize}
\def\babar{\mbox{\slshape B\kern-0.1em{\smaller A}\kern-0.1em
    B\kern-0.1em{\smaller A\kern-0.2em R}}\xspace}

\def\nnn#1#2#3{\hskip -6pt{}^{\phantom{#3}#2}_{\phantom{#2}#3}#1}

\def\etal {{\it et al.}}

\begin{document}

\title{Progress in Lorentz and CPT Violation}

\author{V.\ Alan Kosteleck\'y}

\address{Physics Department, Indiana University\\
Bloomington, IN 47405, USA}

\begin{abstract}
This talk at the CPT'25 meeting presents an overview
of some recent results in Lorentz and CPT violation.
\end{abstract}

\bodymatter

\section{Introduction}

Lorentz transformations,
which include spatial rotations and boosts,
play a central role in our most successful description 
of nature at the fundamental level.
In this description,
the electromagnetic, weak, and strong interactions
are encompassed by the Standard Model (SM),
which is formulated on Minkowski spacetime
and is invariant under global Lorentz transformations
and translations.
Gravitational effects are governed by General Relativity (GR),
which involves Riemann spacetime
and is invariant under local Lorentz transformations
and diffeomorphisms.

An enticing idea that has received substantial attention in recent years
is the notion that violations of Lorentz invariance 
and its concomitant CPT symmetry may be manifest in nature.
These violations could emerge as residual effects
from a unified theory of quantum physics and gravity
at the Planck scale such as strings,\cite{ksp}
and the consequent small but observable signals 
continue to be the subject of
numerous experimental searches.\cite{tables} 
This talk at the {\it Tenth Meeting on CPT and Lorentz Symmetry} (CPT'25)
provides an overview of a few developments in the three years
since the prior meeting in the series.

The treatment adopted here uses effective field theory\cite{eft}
to describe prospective Lorentz- and CPT-violating effects 
in a realistic framework built on GR and the SM,
known as the Standard-Model Extension (SME).\cite{ck,ak04,kl21}
An outline of key concepts in this approach 
can be found in the proceedings of the prior meeting
in this series.\cite{ak-cpt22}
The present contribution summarizes recent applications of this formalism
to the geometry of Finsler spaces associated with Lorentz violation, 
the resolution of the concordance problem
in Lorentz-violating effective field theories,
the relativistic evolution of a spin in background fields,
and the phenomenology of searches 
for flavor-changing Lorentz and CPT violation in charged leptons.

\section{Finsler geometry}

The general Lagrange density describing a Dirac fermion
propagating in a gravitational background
can incorporate terms breaking local Lorentz and diffeomorphism symmetry.
A term of this type involves a field operator 
transforming nontrivially under these symmetries
and contracted with a coefficient
governing the size of the physical effects.\cite{ak04} 
When the coefficient is externally prescribed,
any local Lorentz violation and diffeomorphism violation 
is nondynamical and hence explicit.
This can result in a conflict between the spacetime 
and the particle dynamics if Riemann geometry is assumed,\cite{ak04,rb15}
producing unique experimental signals.\cite{kl21-2}
Adopting Finsler geometry instead is conjectured
to resolve the conflict.\cite{ak04}
Finsler manifolds~\cite{finsler}
provide a natural generalization of Riemann manifolds,
in which the metric depends not only on location
but also on direction in the tangent space.
The conjecture relates the direction dependence in Finsler geometry
to the prescribed coefficients in the Lagrange density.

In the extended Dirac theory,
the analytic continuation of the classical trajectory 
of a relativistic wave packet in a fixed spacetime
is indeed known to correspond to a geodesic in a Finsler geometry
instead of a Riemann one.\cite{ak11}
Many of the specific examples of this correspondence studied to date
are encompassed by the bipartite spaces,
for which the Finsler norm $F$ is defined to be the sum or difference 
of the conventional Riemann norm $\rho$
and an additional seminorm $\sigma$ 
built using a symmetric nonnegative 2-form ${\bf s}$.
Among the set of bipartite spaces 
are the ${\bf a}$ spaces,\cite{ak11}
where ${\bf s}$ is constructed 
using the SME coefficient $a_\mu$ in the Dirac theory.
The ${\bf a}$ spaces are related to Randers geometry,
which underlies the Zermelo problem of navigation 
in a background wind.\cite{zermelo}
Other bipartite spaces obtained from the Dirac theory
include the ${\bf b}$ and ${\bf ab}$ spaces,\cite{ak11}
the ${\bf face}$ spaces,\cite{kr10}
certain $\bf d$,\cite{rs18} $\bf g$,\cite{rs18}
$\bf H$,\cite{krt12} and $\bf k$\cite{ek18} spaces,
and other special geometries.\cite{sma19}
The ${\bf b}$ spaces are of particular interest
in that they arise from the SME coefficient $b_\mu$ 
governing the dominant CPT-odd spin-dependent modifications to the Dirac theory.
The comparatively simple form of ${\bf s}$ for ${\bf b}$ spaces
enables various results to be obtained,\cite{ak11}
including explicit expressions for geometric objects.
Some other physical situations are known to involve ${\bf b}$ geometries,
including a transversely magnetized chain 
and a bead sliding on a wire.\cite{fl15}

A natural question is how to establish whether two given Finsler norms
represent distinct geometries.
Cartan discovered a symmetric torsion 3-tensor ${\bf C}$
that vanishes if and only if 
the Finsler manifold is a Riemann geometry.\cite{ec}
Another symmetric 3-tensor ${\bf M}$,
known as the Matsumoto tensor,\cite{mm}
vanishes when the Finsler manifold is a Randers geometry.
The ${\bf a}$ spaces also have zero ${\bf M}$.\cite{ak11}
Recently,
a symmetric 3-tensor ${\bf S}$ that vanishes for bipartite spaces
has been constructed.\cite{dek25}
The tensor ${\bf S}$ is expressed using the geometric quantities
$F$ and $\rho$ and their derivatives,
including the Cartan torsion ${\bf C}$
and the Finsler metric ${\bf g}$. 
A symmetric 3-tensor ${\bf B}$ that is zero for ${\bf b}$ spaces
and that can be specified in geometric terms 
has also been found.\cite{dek25}
These results provide support for the conjecture 
that all geometries arising from the extended Dirac theory
can be characterized using appropriate symmetric 3-tensors.
Identifying these tensors is an intriguing open problem
that could lead to a classification of the possible geometries
experienced by a Dirac fermion as it propagates
in a fixed gravitational background
in the presence of local Lorentz and diffeomorphism violations.

\section{The concordance problem}

The use of effective field theory to describe low-energy observables 
of Lorentz and CPT violation
that could emerge from an underlying high-energy theory
has parallels in condensed-matter physics. 
Prime examples of this are provided by Dirac and Weyl semimetals.
For instance,
a quasiparticle in a Weyl semimetal 
with band structure exhibiting isolated touching points 
has dynamics described by a generalized Dirac equation
demonstrating experimentally measurable deviations 
from an emergent (3+1)-dimensional 
Lorentz invariance.\cite{semimetal}
It is thus reasonable to hypothesize that 
the comprehensive SME framework can be adopted 
to investigate the properties of general quasiparticle excitations 
in the band structures of Dirac and Weyl materials,
and conversely that established properties in these materials
can be applied to open issues in the SME formalism.\cite{klmss22}

One longstanding issue in the SME approach
has been the physical interpretation
associated with large Lorentz violation.
The difficulty occurs when the coefficients for Lorentz violation
cannot be treated perturbatively in the chosen reference frame,
or when a system with perturbative coefficients
is viewed by a highly boosted observer.
Under these circumstances,
the description of the system can involve puzzling features
such as negative energies.
Finding a prescription that permits a consistent treatment
of large coefficients for Lorentz violation,
including in highly boosted observer frames,
is called the concordance problem.
Traditional analyses of physical Lorentz violation
avoid the concordance problem by fiat,
imposing the requirements
that the coefficients are small
and that boosts are constrained to concordant frames
where perturbative Lorentz violation is maintained.\cite{kl01}
However,
the parallels between the SME formalism
and the treatment of Dirac and Weyl semimetals
offer a novel opportunity to address the concordance problem.\cite{klss25} 
In particular,
no restriction on the magnitude of Lorentz violation 
appears in semimetals even for weak excitations,
which suggests the concordance problem must be resolvable
in the low-energy limit of at least some
some fundamental Lorentz-violating theories.

Following these ideas,
a physical resolution of the concordance problem 
has recently been identified\cite{klss25} 
for the theory of a Dirac fermion propagating in Minkowski spacetime
with a nonzero SME coefficient $b_\mu$ for Lorentz violation.\cite{ck}
In this model,
the concordance problem is associated 
with the appearance of particles and antiparticles
carrying negative energies.
This feature occurs when the coefficient $b_\mu$
has components large compared to the fermion mass $m$
in the chosen observer frame,
or when components of $b_\mu$ that are small in one frame
become large in the frame of a highly boosted observer. 
A resolution of the concordance problem in this case
thus requires a physically viable understanding
of the negative-energy particles and antiparticles. 

Useful insight can be obtained by consideration
of the properties of the corresponding Weyl semimetal,
for which the Fermi surface is established 
via contact with the thermodynamic bath 
associated with the body of the semimetal.
In particular,
the ground state of the system is fixed by the Fermi surface,
below which all the band energy levels are occupied.
Note that this conventional description
implicitly identifies the relevant reference frame
as the rest frame of the bath,
which is appropriate here because
the bath rest frame is physically identical to the semimetal body.

These facts suggest using thermodynamic arguments
to identify the physical ground state of the $b_\mu$ model
via coupling to a generalized bath.
Thermodynamic equilibrium then requires 
the physical vacuum to have zero Fermi energy,
and so all negative-energy states must be filled.
It follows that the ground state of the $b_\mu$ model
has no positive-energy particles and antiparticles, as usual,
but it does contain physical particles and antiparticles 
with negative energies.
This situation is analogous to the physics in the Weyl semimetal,
which includes filled physical levels
with fermions and antifermions lying below the Fermi surface.
In the $b_\mu$ model,
the choice of bath and hence the relevant bath rest frame
must be fixed by physical considerations.
For prospective practical applications in fundamental physics,
a natural choice of bath is the cosmological fluid in the Universe,
which implies that the bath rest frame
is the known cosmological rest frame
associated with the cosmic microwave background. 
This hypothesis could in principle be experimentally tested
if a physical species with large Lorentz violation is found.

Within this physical setup,
the concordance problem for the fundamental particle model 
is resolved in the same way as in the Weyl semimetal.
For example,
in both cases the negative-energy particles and antiparticles
in the ground state have no effect on stability 
because all levels are filled
and hence no energy can be extracted from the ground state
by transitions from positive to negative levels.
The issue of large Lorentz boosts is also resolved
because the bath fixes a physical rest frame,
which implies that no observer boost can change the physics,
instead at most changing the description of the physics
along with the change of observer coordinates.

\section{Extended relativistic BMT equation}

\def\half{{\textstyle{1\over 2}}}

The relativistic Bargmann-Michel-Telegdi (BMT) equation\cite{bmt59}
provides a classical description
of the behavior of a spin-$\half$ particle
in the presence of a homogeneous electromagnetic field.
It encompasses the Heisenberg equation for the quantum spin operator
while allowing 
for the Thomas precession associated with the particle acceleration,
for the Larmor precession induced by an anomalous magnetic moment,
and for the corresponding precession arising from an electric dipole moment.
The BMT equation is relevant in realistic situations
with charged fermions moving
in homogeneous or near-homogeneous electromagnetic fields. 
For instance,
it plays a key role in the analysis of data 
from storage-ring experiments seeking to measure 
the anomalous magnetic moment of the muon~\cite{mu}
or to discover an electric dipole moment
of the muon or the proton.\cite{edm}

A natural question in this context
is whether an extended relativistic BMT equation can be constructed
that incorporates additional types of homogeneous background fields.
These generically carry Lorentz indices,
as a homogeneous background scalar field
primarily acts to shift the fermion mass.
Any homogeneous background with Lorentz indices
can be identified with an SME coefficient for Lorentz violation
that is spacetime constant in the relevant reference frame
and that arises from explicit Lorentz violation.
This line of reasoning indicates that
the general extended relativistic BMT equation
can be associated with the classical behavior of a spin-$\half$ particle
in the presence of explicit Lorentz violation.
Indeed,
the homogeneous electromagnetic field
could also be viewed in this light,
but for applications to searches for new physics
it is convenient to maintain the distinction 
between effects from the electromagnetic field
and those from other prospective backgrounds.

The Lagrange density for a Dirac spinor field 
interacting with a homogeneous electromagnetic field
in the presence of Lorentz violation
contains operators governing both propagation and interaction,
and their explicit form has been established.\cite{ck,kmdk,kl19}
It is thus feasible in principle to construct 
the extended relativistic BMT equation
for any homogeneous background.
In a recent work,\cite{dkv25}
this construction has been achieved 
for the dominant Lorentz-violating corrections,
which arise from propagation effects involving
operators of mass dimensions three and four 
and from electromagnetic couplings involving 
Lorentz-violating operators of mass dimensions five and six.
The resulting extended relativistic BMT equation
is relevant for searches for background physics beyond the SM
in experiments with charged particles moving in electromagnetic fields.
For example,
it plays a role in searches for Lorentz violation
using storage-ring experiments designed to seek electric dipole moments
for the electron, proton, and muon.\cite{dkv25}

\section{Charged-lepton flavor violation}

Flavor-changing physics in the minimal SM
is driven by the weak interactions,
leading to mixing of quark flavors
and transitions from charged leptons to neutrinos.
Beyond the minimal SM,
neutrino oscillations reveal the existence of 
flavor-changing effects with lepton-number violation.
Charged-lepton flavor-violating processes
then must also occur,
although they are one-loop suppressed
with tiny branching ratios $\lesssim10^{-54}$.\cite{smdecays}
This implies that experiments seeking 
charged-lepton flavor-violating processes
can provide sharp tests of new physics.\cite{clfv}

Given this situation,
it is natural to consider the prospects 
for experimental searches for Lorentz and CPT violation 
involving flavor changes in the charged leptons.
Most investigations along these lines concern 
flavor changes during propagation rather than in interactions.
For example,
the dominant effects of Lorentz violation in neutrinos 
involve operators modifying neutrino mixing during propagation.\cite{km12} 
Similarly,
the dominant Lorentz-violating effects 
in neutral-meson oscillations 
arise from modifications to the quark propagators.\cite{ak98}
However,
Lorentz-violating interactions can also induce flavor-changing effects,
notably in charged-lepton flavor-violating processes
of interest here.

All flavor-changing operators of mass dimension $d\leq 6$ 
in the SME Lagrange density 
have been classified and enumerated.\cite{kl19}
Among these are Lorentz-violating operators 
involving electromagnetic interactions with $d=5$,
which contribute already at tree level
to flavor-changing electromagnetic decays 
$\ell_A \rightarrow \ell_B + \gamma$
of a charged lepton $\ell_A$ of flavor $A$ 
into a charged lepton $\ell_B$ of flavor $B\neq A$ 
and a photon.\cite{kps22}
Experimental constraints on the muon branching ratio of
BR$(\mu^+ \rightarrow e^+ +\gamma) \leq 4.2\times 10^{-13}$
have been obtained 
by the Mu to Electron Gamma (MEG) collaboration
at the Paul Scherrer Institute,\cite{MEGlimit}
while limits for the tau of
BR$(\tau^\pm \rightarrow \mu^\pm +\gamma) \leq 4.4\times 10^{-8}$
and 
BR$(\tau^\pm \rightarrow e^\pm +\gamma) \leq 3.3\times 10^{-8}$
have been placed by the \babar collaboration 
at the Stanford Linear Accelerator.\cite{BABARlimit}
These experimental results have been analyzed
to yield limits expressed in the canonical Sun-centered frame\cite{scf}
on coefficients for Lorentz violation 
governing electromagnetic flavor-changing interactions 
in the charged-lepton sector,
lying in the range of parts in 
$10^{-13}$ GeV$^{-1}$ to $10^{-9}$ GeV$^{-1}$.\cite{kps22}
Further analyses of these data
allowing for sidereal and annual time variations 
and other planned experiments\cite{otherexpts}
offer interesting prospects for improved sensitivities.

Another golden channel for investigations
of charged-lepton flavor-violating processes
is the coherent conversion $\mu + N \to e + N$
of a muon into an electron in the presence of a nucleus $N$.
In practice the muon is captured in an atomic ground state,
and the experimental signature
is the monoenergetic spectrum arising from direct conversion.
Tree-level contributions to this process
are induced by Lorentz-violating operators 
involving electromagnetic interactions of mass dimension five
or four-point interactions of leptons and quarks of mass dimension six
in nuclear-mediated processes.\cite{kmps25} 
Using~$\nnn{\rm Au}{197}{79}$ as the target nucleus, 
the SINDRUM~II collaboration at the Paul Scherrer Institute 
has placed the tightest limit to date
on the dimensionless ratio $R_{\mu e}$
of the conversion rate to the capture rate,
$R_{\mu e} < 7\times 10^{-13}$.\cite{sindrum}
Forthcoming searches using~$\nnn{\rm Al}{27}{13}$ as the target
include one by the 
Muon-to-Electron Conversion (Mu2e) collaboration\cite{mu2e}
at Fermilab,
which aims to reach $R_{\mu e} \simeq 6.2\times 10^{-16}$,
and one by the 
Coherent Muon to Electron Transition (COMET) collaboration\cite{comet}
at the Japan Proton Accelerator Complex (J-PARC),
which expects to attain $R_{\mu e} \simeq 7\times 10^{-15}$.
The SINDRUM~II data have recently been analyzed
to yield limits at parts in $10^{-13}$~GeV$^{-2}$
on coefficients for Lorentz violation 
controlling dimension-six flavor-changing four-point interactions
of leptons and quarks,
along with limits at parts in $10^{-12}$~GeV$^{-1}$
on coefficients for Lorentz violation
governing dimension-five flavor-changing 
electromagnetic interactions.\cite{kmps25} 
Future prospects include analyses of existing data
using sidereal and annual variations,
and improved experimental sensitivities are predicted  
in the forthcoming Mu2e and COMET experiments.\cite{kmps25}

\section*{Acknowledgments}

This work was supported in part
by US DoE grant {DE}-SC0010120
and by the Indiana University Center for Spacetime Symmetries.

\end{document}